\documentclass[preprint,showpacs,preprintnumbers,amsmath,amssymb]{revtex4}
\usepackage{graphicx}
\usepackage{dcolumn}
\usepackage{bm}
\usepackage{color}
\usepackage{hhline}

\begin{document}

\title{Arbitrary tip orientation in STM simulations: 3D WKB theory and application to W(110)}

\author{G\'abor M\'andi$^{1}$, Norbert Nagy$^{2}$, and Kriszti\'an Palot\'as$^{1,3,}$}
\email{palotas@phy.bme.hu}
\affiliation{$^{1}$Budapest University of Technology and Economics, Department of Theoretical Physics,
Budafoki \'ut 8., H-1111 Budapest, Hungary\\
$^{2}$Institute for Technical Physics and Materials Science, Research Centre for Natural Sciences,
Hungarian Academy of Sciences, P.\ O.\ Box 49, H-1525 Budapest, Hungary\\
$^{3}$Condensed Matter Research Group of the Hungarian Academy of Sciences,
Budafoki \'ut 8., H-1111 Budapest, Hungary}

\date{\today}

\begin{abstract}

We extend the orbital-dependent electron tunneling model implemented within the three-dimensional (3D) Wentzel-Kramers-Brillouin
(WKB) atom-superposition approach for simulating scanning tunneling microscopy (STM) by including arbitrary tip orientations.
The orientation of the tip is characterized by a local coordinate system centered on the tip apex atom obtained by a rotation
with respect to the sample coordinate system. The rotation is described by the Euler angles.
Applying our method, we highlight the role of the real-space shape of the electron orbitals involved in the tunneling,
and analyze the convergence and the orbital contributions of the tunneling current above the W(110) surface depending
on the orientation of a model tungsten tip. We also simulate STM images at constant-current condition, and find that their quality
depends very much on the tip orientation. Some orientations result in protrusions on the images that do not occur above W atoms.
The presence of such apparent atom positions makes it difficult to identify the exact position of surface atoms. It is suggested
that this tip orientation effect should be considered at the evaluation of experimental STM images on other surfaces as well.
The presented computationally efficient tunneling model could prove to be useful for obtaining more information on the local tip
geometry and orientation by comparing STM experiments to a large number of simulations with systematically varied
tip orientations.

\end{abstract}


\maketitle

\section{Introduction}

The scanning tunneling microscope (STM) is undoubtedly a successfully used tool to study physical and chemical phenomena on
surfaces of materials. The experimentally least controllable parameter of the STM equipment is the local geometry and orientation
of the tip structure, that plays an ultimate role in determining the electron tunneling features
\cite{cerda97,hofer03rmp,hofer03pssci,paz05,hagelaar08}. The problem of the presence of multiple tip apices or nanotips can also
arise \cite{rodary11}. Therefore advanced theoretical models are needed that are capable to deal with large scale simulations of a
considerable variety of realistic tip structures \cite{hagelaar08}. The present work tries to complement existing
methods, and introduces a computationally efficient model based on the orbital-dependent atom-superposition tunneling approach
\cite{palotas12orb} that allows a large flexibility of tip structures, and orientations in particular.

Owing to the practically unknown tip structure, the identification of atomic positions from experimentally observed STM images is
not straightforward. The tip size effect on asymmetric surface features has been demonstrated to result in
the occurrence of an apparent step edge shifted away from the real geometric position on a Au(11 12 12) surface \cite{xiao06}.
To model the tip size and shape effect, an $s$-orbital continuum model has been proposed \cite{gaspari11}.
Tip rotations within the extended H\"uckel theory have also been put forward \cite{hagelaar08}. All of these listed examples
studied highly corrugated surfaces having topographic features of at least one atom height difference on top of a flat surface.

It is interesting to find that identifying the atomic positions even on flat surfaces can be problematic. The reason is
the corrugation inversion phenomenon found, e.g., on (100) \cite{mingo96}, (110) \cite{heinze98}, and (111) \cite{ondracek12}
metal surfaces. According to Heinze {\it {et al.}} \cite{heinze98}, it was found on a W(110) surface that under certain
circumstances the apparent height of tungsten atoms at the surface top position can be larger or smaller than the
apparent height of the surface hollow position at constant-current condition. It means that metal atoms do not
always appear as protrusions on the STM image.
It was reported that the W(110) surface has a corrugation inversion depending on the bias voltage \cite{heinze98}, and on the
tip-sample distance and tip orbital character as well \cite{palotas12orb}. Chen explained this effect occurring on low Miller
index metal surfaces as a consequence of $m\ne 0$ tip states \cite{chen92}.
Atomic contrast reversal has also been found above Xe atomic adsorbates \cite{mingo96}
and oxygen overlayers \cite{calleja04} on metal surfaces. It was established that the character of the contrast depends on the
tip-sample distance and on the tip geometry and electronic structure.

The effect of the tip on the electron tunneling properties has been studied in numerous works. For example,
Ness and Gautier investigated different metal tips and their interaction with metal surfaces in a tight-binding framework
\cite{ness95jpcm1,ness95jpcm2,ness95prb}.
Ref.\ \cite{hofer05sts} presented a theoretical method that can separate the tip and sample contributions to the $dI/dV$ in
scanning tunneling spectroscopy (STS).
The difference between model magnetic tips on the spin-polarized spectroscopic properties (SP-STS) was investigated in
Refs.\ \cite{palotas12sts,donati13}. Magnetic contrast variations were studied in spin-polarized STM (SP-STM)
on a noncollinear magnetic surface in Refs.\ \cite{palotas11stm,palotas13contrast}.
Teobaldi {\it {et al.}} rationalized the STM contrast mechanisms observed on the graphite(0001) surface by modeling a few
tungsten tips taking the effects of tip termination, composition, and sharpness into account \cite{teobaldi12}.

Different electron transport models considered the role of the electron orbitals.
Chen and Sacks theoretically studied the effect of the tip orbitals
on the corrugation of constant-current STM images \cite{chen90,sacks00}.
While Chen pointed out that corrugation enhancement is expected for tip orbitals localized along the surface normal ($z$)
direction ($p_z$ and $d_{3z^2-r^2}$), Sacks argued that $m\ne 0$ tip states ($d_{xz}$, $d_{yz}$, $d_{xy}$, $d_{x^2-y^2}$) are
responsible for this effect.
Sirvent {\it {et al.}} presented a tight-binding model based on the Keldysh formalism for calculating the conductance in atomic
point contacts and analyzed the effect of the $d$ orbitals \cite{sirvent96}.
Mingo {\it {et al.}} used the same method for STM junctions \cite{mingo96}.
Cerd\'a {\it {et al.}} developed an STM simulation method based on the Landauer-B\"uttiker formula \cite{buttiker85} and the
surface Green function matching technique \cite{cerda97}.
Palot\'as {\it {et al.}} introduced an orbital-dependent tunneling transmission function within an atom-superposition
framework \cite{palotas12orb}.
In these methods the decomposition of the current and/or conductance with respect to electron orbitals has been provided.

In the present work we consider a simple model for orbital-dependent tunneling within the atom-superposition approach based on
Ref.\ \cite{palotas12orb}. The main idea of the paper is the extension of the
geometrical factor, responsible for a modified transmission due to electron orbital orientational overlap effects, to an arbitrary
local tip coordinate system within a three-dimensional (3D) Wentzel-Kramers-Brillouin (WKB)-based theory \cite{palotas13fop}.
The method is used for investigating the nature of apparent atomic positions on STM images depending on the tip orientation.
We provide a basic understanding of the features based on the real-space shape of the electron orbitals involved in the tunneling.

The paper is organized as follows: The theoretical model of the orbital-dependent tunneling within the atom-superposition
approach taking arbitrary tip orientations into account is presented in section \ref{sec_theory}. We investigate the convergence
and the orbital contributions of the tunneling current as well as the atomic contrast changes of the W(110) surface depending on
the orientation of a model tungsten tip in section \ref{sec_res}. Summary of our findings is found in section \ref{sec_conc}.

\section{Orbital-dependent tunneling model within 3D WKB theory with arbitrary tip orientation}
\label{sec_theory}

Palot\'as {\it{et al.}} developed an orbital-dependent electron tunneling model \cite{palotas12orb} for simulating STM and STS
measurements within the 3D WKB framework based on previous atom-superposition theories \cite{tersoff85,yang02,smith04,heinze06}
and an STS theory \cite{passoni09}. Using this method provided comparable STM images to those obtained by standard
Tersoff-Hamann \cite{tersoff83,tersoff85} and Bardeen \cite{bardeen61} tunneling models implemented in the BSKAN code
\cite{hofer03pssci,palotas05}, and has successfully been used for re-investigating the corrugation inversion phenomenon on the
W(110) surface \cite{heinze98,palotas12orb}. The advantages, particularly computational efficiency, limitations, and the potential
of the method have been discussed in Ref.\ \cite{palotas13fop}. Here, we extend this model by considering an arbitrary tip
orientation. In the model, it is assumed that electrons tunnel through one tip apex atom only, and for the tunneling process the
one-dimensional (1D) WKB approximation is used. Tunneling transitions between the tip apex atom and a suitable number of sample
surface atoms are summed up \cite{palotas11sts,palotas12orb}. Since the 3D geometry of the tunnel junction is considered, the
method is, in effect, a 3D WKB approach. Another 3D approach for STS with a prescribed tip orbital symmetry has been reported by
Donati {\it{et al.}} \cite{donati11}. In our method, the electronic structure of the tip and the surface is included via the
atom-projected electron density of states (PDOS) obtained by {\it{ab initio}} electronic structure calculations
\cite{palotas11stm}. The orbital-decomposition of the PDOS is essential for the description of the orbital-dependent tunneling
\cite{palotas12orb}.

Assuming elastic tunneling and $T=0$ K temperature, the tunneling current measured at $\mathbf{R}_{TIP}$ tip position with $V$
bias voltage is given as
\begin{equation}
\label{eq:current}
I\left(\mathbf{R}_{TIP},V\right)=\int_0^V\frac{dI}{dU}\left(\mathbf{R}_{TIP},U,V\right)dU.
\end{equation}
The integrand can be written as a superposition of individual atomic contributions from the sample surface (sum over $a$):
\begin{eqnarray}
\label{eq:dIdU}
&&\frac{dI}{dU}\left(\mathbf{R}_{TIP},U,V\right)\\
&=&\varepsilon^{2}\frac{e^{2}}{h}\sum_a\sum_{\beta,\gamma}T_{\beta\gamma}\left(E_F^S+eU,V,\mathbf{d}_{a}\right)n_{S\beta}^{a}\left(E_F^S+eU\right)n_{T\gamma}\left(E_F^T+eU-eV\right).\nonumber
\end{eqnarray}
Here, $e$ is the elementary charge, $h$ is the Planck constant, and $E_F^S$ and $E_F^T$ are the Fermi energies of the
sample and the tip, respectively. The $\varepsilon^{2}e^{2}/h$ factor gives the correct dimension (A/V) of the formal
conductance-like quantity in Eq.(\ref{eq:dIdU}). The value of $\varepsilon$ has to be determined by comparing the simulation
results to experiments, or to calculations with other methods, e.g., the Bardeen approach \cite{bardeen61}. In our simulations
$\varepsilon=1$ eV has been chosen, that gives comparable current values to the ones obtained from the Bardeen method
\cite{palotas12orb}. Note that the choice of $\varepsilon$ has no qualitative influence on the reported results.
$n_{S\beta}^{a}\left(E\right)$ and $n_{T\gamma}\left(E\right)$ are the orbital-decomposed PDOS functions for the $a$th sample
surface atom and the tip apex atom with orbital symmetry $\beta$ and $\gamma$, respectively. These quantities can be obtained
by any suitable electronic structure calculation. The total PDOS is simply the sum of the orbital-decomposed contributions:
\begin{eqnarray}
n_{S}^{a}\left(E\right)=\sum_{\beta}n_{S\beta}^{a}\left(E\right),\\
n_{T}\left(E\right)=\sum_{\gamma}n_{T\gamma}\left(E\right).
\end{eqnarray}
Note that a similar decomposition of the Green functions was used within the linear combination of atomic orbitals (LCAO)
framework in Refs.\ \cite{mingo96,sirvent96}.

The sum over $\beta$ and $\gamma$ in Eq.(\ref{eq:dIdU}) denotes the superposition of the effect of atomic orbitals of the sample
and the tip, respectively, via an orbital-dependent tunneling transmission function:
$T_{\beta\gamma}\left(E_F^S+eU,V,\mathbf{d}_{a}\right)$ gives the
probability of the electron tunneling from the $\beta$ orbital of the $a$th surface atom to the $\gamma$ orbital of the tip apex
atom at negative bias voltage ($V<0$), and from the tip to the surface at positive bias ($V>0$). The transmission probability
depends on the energy of the electron (measured from the sample Fermi energy), the bias voltage ($V$), and the relative position
of the tip apex and the $a$th sample atom ($\mathbf{d}_{a}=\mathbf{R}_{TIP}-\mathbf{R}_{a}$). In our model, we consider
$\beta,\gamma\in\{s,p_y,p_z,p_x,d_{xy},d_{yz},d_{3z^2-r^2},d_{xz},d_{x^2-y^2}\}$ atomic orbitals,
and the following form for the transmission function:
\begin{equation}
\label{Eq_Transmission_decomp}
T_{\beta\gamma}\left(E_F^S+eU,V,\mathbf{d}_{a}\right)=e^{-2\kappa(U,V)d_{a}}t_{\beta\gamma}\left(\vartheta_{a},\varphi_{a},\vartheta_{a}',\varphi_{a}'\right).
\end{equation}
The exponential factor corresponds to an orbital-independent transmission, where all electron states are considered to be
exponentially decaying spherical states \cite{tersoff83,tersoff85,heinze06}, and it depends on the distance between the $a$th
surface atom and the tip apex, $d_{a}=|\mathbf{d}_{a}|$, and on the vacuum decay,
\begin{equation}
\kappa(U,V)=\frac{1}{\hbar}\sqrt{2m\left(\frac{\phi_{S}+\phi_{T}+eV}{2}-eU\right)}.
\end{equation}
Here, we assumed an effective rectangular potential barrier in the vacuum between the sample and the tip. $\phi_{S}$ and
$\phi_{T}$ are the electron work functions of the sample and the tip, respectively, $m$ is the electron mass, and $\hbar$ is the
reduced Planck constant. The method of determining the electron work functions from the calculated local electrostatic potential
is reported, e.g., in Ref.\ \cite{palotas11stm}.

\begin{figure*}
\includegraphics[width=0.75\textwidth,angle=0]{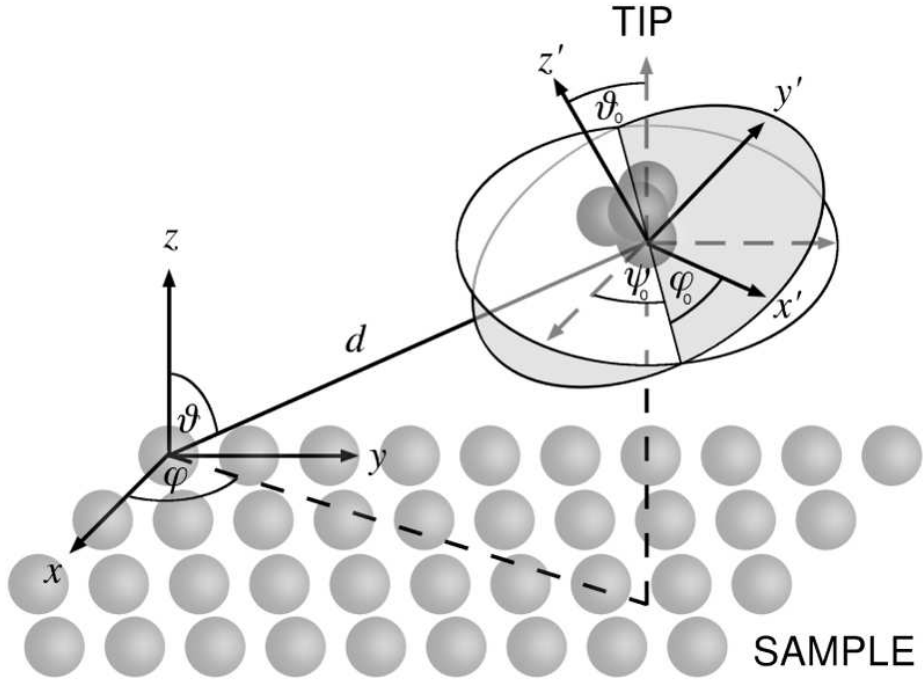}
\caption{\label{Fig1} Geometry of a general tip-sample setup. The rotation of the tip coordinate system is described by the
Euler angles $(\vartheta_0,\varphi_0,\psi_0)$.
}
\end{figure*}

The orbital-dependence of the transmission
coefficient is given by the geometry factor $t_{\beta\gamma}\left(\vartheta_{a},\varphi_{a},\vartheta_{a}',\varphi_{a}'\right)$
that takes into account the relative orientation of the real-space shape of different atomic orbitals between the $a$th sample
atom and the tip apex atom. The definition of this factor has the following physical motivation: It modifies the exponentially
decaying transmission probability according to the angular dependence of the electron densities of the atomic orbitals.
The concept is discussed in more detail in Ref.\ \cite{palotas12orb}. The angular dependence of an atomic orbital of a
sample (tip) atom is given by the real spherical harmonics $\chi_{\beta(\gamma)}(\vartheta^{(}$$'$$^{)},\varphi^{(}$$'$$^{)})$
that is defined in the local coordinate system fixed to the sample surface atom $\left(r,\vartheta,\varphi\right)$,
or to the tip apex atom $\left(r',\vartheta',\varphi'\right)$, and depends on the local polar ($\vartheta^{(}$$'$$^{)}$) and
azimuthal ($\varphi^{(}$$'$$^{)}$) angles. An arbitrary tip orientation corresponds to a rotated local tip coordinate system
with respect to the coordinate system chosen for the $a$th surface atom, see Fig.\ \ref{Fig1}. Generally, we have to distinguish
between these two coordinate systems, so that the coordinates of a given vector are denoted with primes
($'$) if they are defined in the rotated coordinate system of the tip: $(x',y',z')$, and without primes if defined in the
coordinate system of the sample $(x,y,z)$. The rotation of the axes with respect to each other is given by the Euler angles
$(\vartheta_0,\varphi_0,\psi_0)$ shown in Fig.\ \ref{Fig1}:
\begin{equation}
\left[
\begin{array}{c}
x'\\
y'\\
z'
\end{array}
\right]
=\underline{\underline{R}}(\vartheta_0,\varphi_0,\psi_0)
\left[
\begin{array}{c}
x\\
y\\
z
\end{array}
\right],
\end{equation}
and the rotation matrix is
\begin{eqnarray}
\label{Eq_mrot}
&&\underline{\underline{R}}(\vartheta_0,\varphi_0,\psi_0)=\\
&&
\left[
\begin{array}{ccc}
\cos\varphi_0\cos\psi_0-\sin\varphi_0\sin\psi_0\cos\vartheta_0 & \cos\varphi_0\sin\psi_0+\sin\varphi_0\cos\psi_0\cos\vartheta_0 & \sin\varphi_0\sin\vartheta_0\\
-\sin\varphi_0\cos\psi_0-\cos\varphi_0\sin\psi_0\cos\vartheta_0 & -\sin\varphi_0\sin\psi_0+\cos\varphi_0\cos\psi_0\cos\vartheta_0 & \cos\varphi_0\sin\vartheta_0\\
\sin\psi_0\sin\vartheta_0 & -\cos\psi_0\sin\vartheta_0 & \cos\vartheta_0
\end{array}
\right].\nonumber
\end{eqnarray}

With these physical and geometrical considerations, the orbital-dependent part of the transmission probability is the following:
\begin{equation}
t_{\beta\gamma}\left(\vartheta_{a},\varphi_{a},\vartheta_{a}',\varphi_{a}'\right)=\chi_{\beta}^{2}\left(\vartheta_{a},\varphi_{a}\right)\chi_{\gamma}^{2}\left(\vartheta_{a}',\varphi_{a}'\right).
\end{equation}
Here $\chi_{\beta}\left(\vartheta_{a},\varphi_{a}\right)$ and $\chi_{\gamma}\left(\vartheta_{a}',\varphi_{a}'\right)$ give the
angular dependence of the electron wave function of the $\beta$ orbital of the $a$th surface atom and of the $\gamma$ orbital
of the tip apex atom, respectively, see Table \ref{Table1}. The squares of these functions give the angular dependence of the
corresponding electron densities. The angles given in both real spherical harmonics correspond to the
tunneling direction, i.e., the line connecting the $a$th surface atom and the tip apex atom, as viewed from their local coordinate
systems. If the geometrical positions of the tip and the surface atoms are given in global coordinates then the angles
$\left(\vartheta_{a},\varphi_{a}\right)$ can be obtained from the following equations:
\begin{eqnarray}
\mathbf{d}_{a}&=&\mathbf{R}_{TIP}\left(x,y,z\right)-\mathbf{R}_{a}\left(x_{a},y_{a},z_{a}\right)\nonumber\\
&=&\left(x-x_{a},y-y_{a},z-z_{a}\right)=\left(d_{a},\vartheta_{a},\varphi_{a}\right),\\
d_{a}&=&\sqrt{\left(x-x_{a}\right)^{2}+\left(y-y_{a}\right)^{2}+\left(z-z_{a}\right)^{2}},\\
\vartheta_{a}&=&\arccos\left(\frac{z-z_{a}}{d_{a}}\right),\\
\varphi_{a}&=&\arccos\left(\frac{x-x_{a}}{d_{a}\sin\vartheta_{a}}\right).
\end{eqnarray}
Similarly, the angles $\left(\vartheta_{a}',\varphi_{a}'\right)$ can be calculated by expressing the coordinates of the vector
$-\mathbf{d}_{a}$ in the local coordinate system of the tip, as follows:
\begin{eqnarray}
\mathbf{d}_{a}'&=&-\underline{\underline{R}}(\vartheta_0,\varphi_0,\psi_0)\mathbf{d}_{a}=\left(x_{a}',y_{a}',z_{a}'\right)=\left(d_{a}',\vartheta_{a}',\varphi_{a}'\right),\\
d_{a}'&=&\sqrt{x_{a}'^{2}+y_{a}'^{2}+z_{a}'^{2}}=d_{a},\\
\vartheta_{a}'&=&\arccos\left(\frac{z_{a}'}{d_{a}}\right),\\
\varphi_{a}'&=&\arccos\left(\frac{x_{a}'}{d_{a}\sin\vartheta_{a}'}\right).
\end{eqnarray}
Here, $\underline{\underline{R}}(\vartheta_0,\varphi_0,\psi_0)$ is defined in Eq.(\ref{Eq_mrot}).

From Eqs.(\ref{eq:current}) and (\ref{eq:dIdU}) it is clear that the tunneling current can be decomposed according to the orbital
symmetries:
\begin{equation}
I\left(\mathbf{R}_{TIP},V\right)=\sum_{\beta,\gamma}I_{\beta\gamma}\left(\mathbf{R}_{TIP},V\right),
\end{equation}
with
\begin{equation}
I_{\beta\gamma}\left(\mathbf{R}_{TIP},V\right)=\int_0^V\frac{dI_{\beta\gamma}}{dU}\left(\mathbf{R}_{TIP},U,V\right)dU,
\end{equation}
and
\begin{eqnarray}
\label{eq:I_full_decomp}
&&\frac{dI_{\beta\gamma}}{dU}\left(\mathbf{R}_{TIP},U,V\right)\\
&=&\varepsilon^{2}\frac{e^{2}}{h}\sum_a T_{\beta\gamma}\left(E_{F}^{S}+eU,V,\mathbf{d}_{a}\right)n_{S\beta}^{a}\left(E_{F}^{S}+eU\right)n_{T\gamma}\left(E_{F}^{T}+eU-eV\right).\nonumber
\end{eqnarray}
This decomposition gives the opportunity to analyze the tunneling process in terms of orbital contributions.
The relative contribution of the $\beta\leftrightarrow\gamma$ transition can be calculated as
\begin{equation}
\label{Eq_i_bg}
\tilde{I}_{\beta\gamma}\left(\mathbf{R}_{TIP},V\right)=\frac{I_{\beta\gamma}\left(\mathbf{R}_{TIP},V\right)}{I\left(\mathbf{R}_{TIP},V\right)}.
\end{equation}

Using the presented method, we can investigate tip rotational effects on the tunneling properties, e.g., on the STM image.
This could prove to be extremely useful if one wants to gain information on the local geometrical properties of the tip in
real STM experiments by comparing measurements to simulation results \cite{hagelaar08}.

\section{Results and discussion}
\label{sec_res}

To demonstrate the tip orientation effects on the tunneling properties we consider a W(110) surface. This surface is a
widely used substrate for thin film growth, see e.g., Refs.\ \cite{heinze98,bode07}, therefore it has a technological importance.
Heinze {\it {et al.}} \cite{heinze98} pointed out that the determination of the position of surface atomic sites is not
straightforward as atomic resolution is lost at negative bias voltages, and a bias-dependent contrast reversal has been
predicted. This means that normal and anticorrugated constant-current STM images can be obtained in certain bias voltage ranges,
and the W atoms do not always appear as protrusions in the images. It was shown that a competition between states from different
parts of the surface Brillouin zone is responsible for this effect \cite{heinze98,heinze99}. Explanation of this effect based on
the real-space shape of the electron orbitals within an orbital-dependent tunneling model was given in Ref.\ \cite{palotas12orb}.
For an $s$-type tip, an excellent agreement has been found with the results of Ref.\ \cite{heinze98}. Concerning tips with
$p_z$ and $d_{3z^2-r^2}$ orbital symmetry, it was reported in Ref.\ \cite{palotas12orb} that the contrast inversion occurs
at larger tip-sample distances, in contrast to the speculations of Ref.\ \cite{heinze98}. Moreover, it was shown that
two qualitatively different corrugation inversion behaviors can occur based on the tip orbital composition \cite{palotas12orb}.
In the present work, we investigate the atomic contrast changes depending on the tip orientation of a model tungsten tip.

\subsection{Computational details}
\label{sec_comput}

We performed geometry relaxation and electronic structure calculations based on the density functional theory (DFT)
within the generalized gradient approximation (GGA) implemented in the Vienna Ab-initio Simulation Package (VASP)
\cite{VASP2,VASP3,hafner08}. A plane-wave basis set for the electronic wave function expansion, and the
projector-augmented wave (PAW) method \cite{kresse99} for the description of the electron-ion interaction were employed.
We used the Perdew-Wang (PW91) parametrization \cite{pw91} of the exchange-correlation functional.
The electronic structures of the sample surface and the tip were calculated separately.

\begin{figure*}
\includegraphics[width=0.40\textwidth,angle=0]{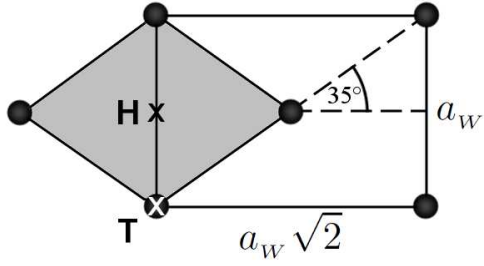}
\caption{\label{Fig2} The surface unit cell of W(110) (shaded area) and the rectangular scan area for the tunneling current
simulations. Circles denote the W atoms. The top (T) and hollow (H) positions are explicitly shown.
}
\end{figure*}

We modeled the W(110) surface by a slab of nine layers, where the two topmost W layers have been fully relaxed.
We used the experimental lattice constant of $a_W=316.52$ pm.
The unit cell of the W(110) surface (shaded area), the rectangular scan area for the tunneling current simulation, and the
surface top (T) and hollow (H) positions are shown in Figure \ref{Fig2}.
A $41\times 41\times 5$ Monkhorst-Pack (MP) \cite{monkhorst} k-point grid was used for obtaining the orbital-decomposed
projected electron DOS onto the surface W atom, $n_{S\beta}^{a}(E)$.

Motivated by a previous work \cite{teobaldi12}, we considered a blunt W(110) tip model, i.e., an adatom adsorbed on the hollow
site as the tip apex on the W(110) surface. The adatom position has been relaxed in the surface normal direction.
Moreover, an $11\times 15\times 5$ MP k-point grid was used for calculating the orbital-decomposed projected DOS onto
the apex atom, $n_{T\gamma}(E)$. The electron work functions of the sample and the tip were chosen as $\phi_S=\phi_T=4.8$ eV.
More details about the relaxed surface and tip structures can be found in Ref.\ \cite{palotas12orb}.

Using the presented model, the following tip orientations were calculated:
$\vartheta_0\in[0^{\circ},75^{\circ}]$ and $\varphi_0,\psi_0\in[0^{\circ},90^{\circ}]$ with $5^{\circ}$ steps.
We report selected results of this big data set highlighting the tip orientation trends on the tunneling properties.
We consider the following sets for the Euler angles $(\vartheta_0,\varphi_0,\psi_0)$:
$(0^{\circ},0^{\circ},[0^{\circ},90^{\circ}])$,
$([0^{\circ},75^{\circ}],0^{\circ},0^{\circ})$,
$(45^{\circ},0^{\circ},[0^{\circ},90^{\circ}])$,
$(45^{\circ},[0^{\circ},90^{\circ}],0^{\circ})$.
Note that by changing the Euler angles, tunneling through one tip apex atom was considered only, and contributions from other
tip atoms were not taken into account. High degrees of tilting the tip ($\vartheta_0>45^{\circ}$) could, in fact,
result in multiple tip apices or nanotips \cite{rodary11} depending on the local geometry that can increase the tunneling current
but could also lead to the destruction of atomic resolution.

The tunneling current was calculated in a box above the rectangular scan area shown in Figure \ref{Fig2}
containing 99000 ($30\times 22\times 150$) grid points with a $0.149$ \AA\;lateral and $0.053$ \AA\;vertical resolution.
The constant-current contours are extracted following the method described in Ref.\ \cite{palotas11stm}, and
we report STM images above the mentioned rectangular scan area.

\subsection{Convergence properties}
\label{sec_conv}

\begin{figure*}
\includegraphics[width=0.80\textwidth,angle=0]{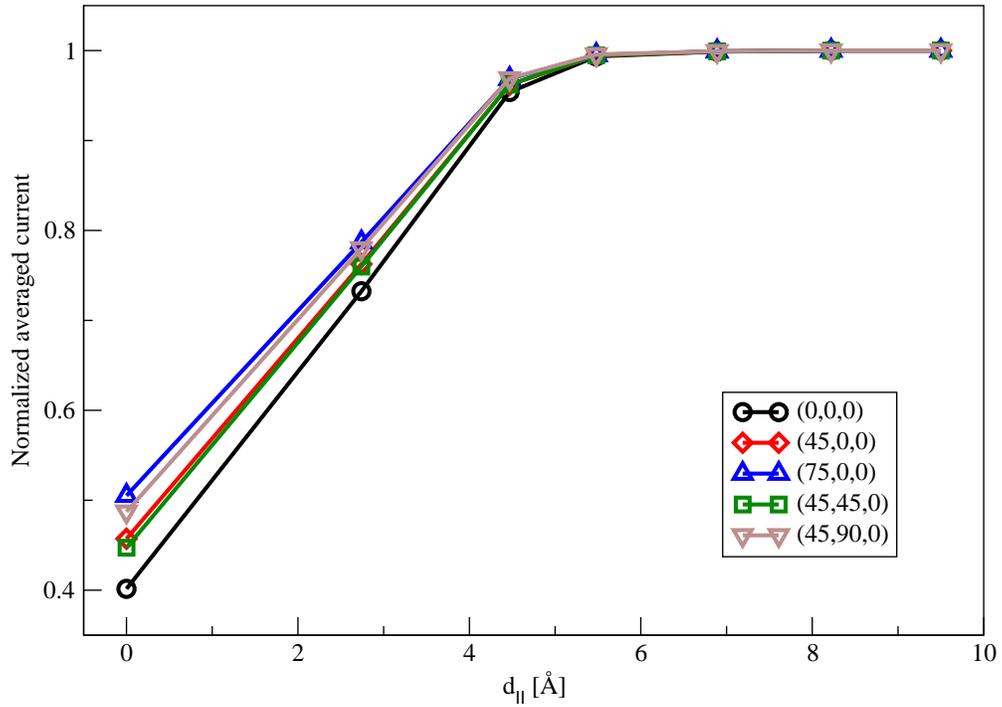}
\caption{\label{Fig3} (Color online) Convergence of the normalized averaged current $z$=4.5 \AA\; above the surface top (T)
position (W atom) calculated with different tungsten tip orientations described by the Euler angles
$(\vartheta_0,\varphi_0,\psi_0)$ given in degrees, see also Figure \ref{Fig1}.
}
\end{figure*}

Previously, the convergence of the tunneling current was investigated with respect to the number of surface atoms involved
in the summation of the atom-superposition formula (sum over $a$) without tip rotation \cite{palotas12orb}.
It was found that the orbital-independent, the $s$-type, and the tungsten tips behave similarly concerning the current
convergence, and for the $p_z$- and $d_{3z^2-r^2}$-type tips a faster convergence was found. The latter finding was explained by
the more localized character of the corresponding tip orbitals in the direction normal to the sample surface.
We report a similar convergence test for the tungsten tip comparing different tip orientations.
To take into account a wide energy range around the Fermi level, we calculated the tunneling current at
-2.0 V and +2.0 V bias voltages at $z=4.5$ \AA\; above a surface W atom, and averaged these current values.
The averaged currents were normalized for each tip calculation to obtain comparable results.
The convergences of the normalized averaged current with respect to the lateral distance on the surface, $d_{\parallel}$,
characteristic for the number of atoms involved in the atom-superposition summation, are shown in Figure \ref{Fig3}.
$d_{\parallel}$ represents the radius of a surface section measured from the W atom below the tip apex, from which area the
surface atomic contributions to the tunneling current are taken.

We find that by fixing the $z'=z$ axis ($\vartheta_0=0^{\circ}$), the rotation of the tip with $\psi_0\in[0^{\circ},90^{\circ}]$
does not change the convergence character compared to $\psi_0=0^{\circ}$ (not shown). This is due to the dominant current
contributions from the $s$, $p_z$, and $d_{3z^2-r^2}$ orbitals of both the sample and the tip, that do not change upon the
mentioned tip rotation. For an illustration see the top left part of Figure \ref{Fig4}.
The situation is remarkably different by changing $\vartheta_0$. This tip rotation has an effect of a tilted $z'$
axis of the tip apex compared to the sample $z$ direction. The more the tilting the faster convergence of the normalized averaged
current is observed. We show examples of $(45^{\circ},0^{\circ},0^{\circ})$ and $(75^{\circ},0^{\circ},0^{\circ})$
in Figure \ref{Fig3}. As the rotation of $\vartheta_0$ is around the $x$ axis, i.e., $x'=x$ remains the same, the tip $d_{y'z'}$
and $d_{x'^2-y'^2}$ orbitals with nodal planes involving the $z'$ direction gain more importance in the tunneling as the tilting
increases since they can hybridize easier with the dominant orbitals of the sample: $s$, $p_z$, and $d_{3z^2-r^2}$. This finding
is demonstrated in the top right part of Figure \ref{Fig4}. Concomitantly, the tip $p_{z'}$ and $d_{3z'^2-r'^2}$ orbitals
lose contribution as they give transmission maximum in the $z'$ direction that is not in-line with $z$ because of the tilting.
Starting from the $(45^{\circ},0^{\circ},0^{\circ})$ tip orientation, we can rotate the tip around the sample $z$ direction
with angles $\psi_0\in[0^{\circ},90^{\circ}]$. We find that this type of rotation does not considerably affect the convergence
character of the current compared to the $(45^{\circ},0^{\circ},0^{\circ})$ orientation (not shown). This is due to the
practically unchanged dominant current contributions by rotating with $\psi_0$, see the bottom left part of Figure \ref{Fig4}.
On the other hand, rotating the local tip coordinate system around $z'$, i.e., by changing $\varphi_0$ results in slight
convergence changes. First, the convergence speed drops slightly at $(45^{\circ},45^{\circ},0^{\circ})$, and then increases at
$(45^{\circ},90^{\circ},0^{\circ})$ orientation. This effect is related to the tip $d_{x'z'}$ and $d_{y'z'}$ orbitals
as their contribution changes the most by this type of rotation, see also the bottom right part of Figure \ref{Fig4}.

We found that the tip rotation effects do not change the suggestion that atom contributions within at least
$d_{\parallel}=3a_W\approx 9.5$ \AA\; distance from the surface-projected tip position have to be considered \cite{palotas12orb}.
The reason is that the exponentially decaying part of the transmission function is dominant over the orbital-dependent part.
In case of calculating STM images, $d_{\parallel}=3a_W\approx 9.5$ \AA\; has to be measured from the edge of the scan area in all
directions to avoid distortion of the image, thus involving 67 surface atoms in the atomic superposition.
For brevity, in the following we use the same surface atoms to calculate single-point tunneling properties as well.

\subsection{Orbital Contributions}
\label{sec_contrib}

\begin{figure*}
\includegraphics[width=1.00\textwidth,angle=0]{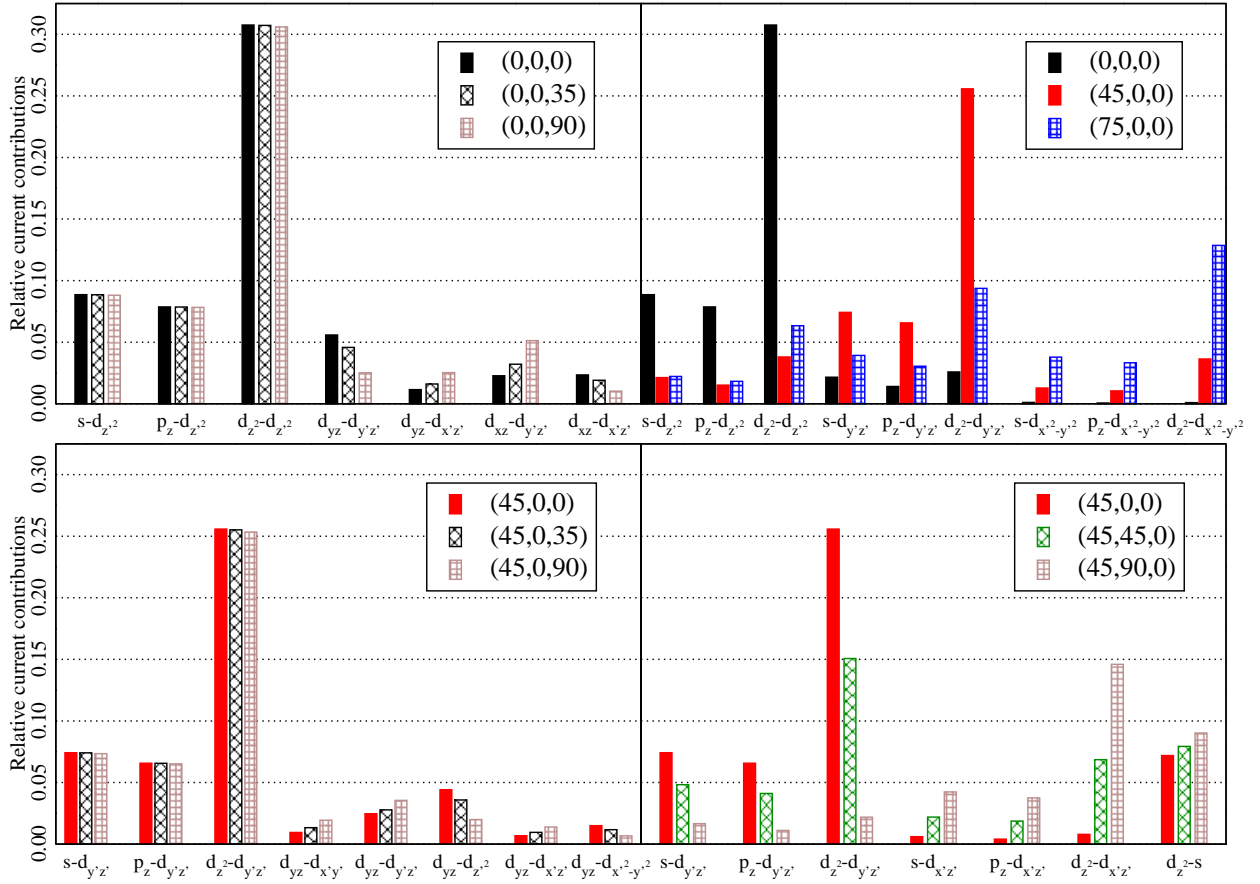}
\caption{\label{Fig4} (Color online) Tip orientation effect on selected relative current contributions between sample $\beta$ and
tip $\gamma$ orbitals [$\tilde{I}_{\beta\gamma}$ in Eq.(\ref{Eq_i_bg}), here denoted by $\beta-\gamma$] using the tungsten tip at
$V$= -0.1 V bias voltage, $z$=4.5 \AA\; above the surface top (T) position (W atom). The tip orientation is described by the
Euler angles $(\vartheta_0,\varphi_0,\psi_0)$ given in degrees, see also Figure \ref{Fig1}.
For brevity, we used the notation $d_{z^{(')2}}$ for the $d_{3z^{(')2}-r^{(')2}}$ orbitals.
}
\end{figure*}

Let us analyze the tip orientation effects on the relative importance of selected $\beta\leftrightarrow\gamma$ transitions in
determining the total tunneling current above a surface W atom. From this analysis we obtain a quantitative picture about the
role of the different atomic orbitals in the construction of the tunneling current, and their changes upon tip rotation.
The $\tilde{I}_{\beta\gamma}$ relative current contributions can be calculated according to Eq.(\ref{Eq_i_bg}). This quantity
gives the percentual contribution of the individual transition to the total tunneling current. Figure \ref{Fig4} shows
selected relative current contributions using the tungsten tip at $V$= -0.1 V bias voltage $z=4.5$ \AA\; above a surface W atom.
Note that those transitions are reported only, which have either a significant contribution, or show considerable changes
upon the tip rotations.
We find that by rotating the tip with $\psi_0$ around the $z'=z$ axis (top left part of Figure \ref{Fig4}),
the dominant contributions are due to the tip $d_{3z'^2-r'^2}$ orbital combined with the sample $s$, $p_z$, and $d_{3z^2-r^2}$
orbitals, and they do not change by the mentioned tip rotation. On the other hand, the $d_{yz}-d_{y'z'}$ and $d_{xz}-d_{x'z'}$
contributions lose, while the $d_{yz}-d_{x'z'}$ and $d_{xz}-d_{y'z'}$ gain importance upon this type of tip rotation.
The top right part of Figure \ref{Fig4} corresponds to rotations around the $x'=x$ axis with $\vartheta_0$, and the evolution of
the dominant contributions. It can be seen that the dominant sample contributions remain unchanged, i.e., they are the
$s$, $p_z$, and $d_{3z^2-r^2}$ orbitals, while the dominating tip orbitals change from $d_{3z'^2-r'^2}$ at
$(0^{\circ},0^{\circ},0^{\circ})$ to $d_{y'z'}$ at $(45^{\circ},0^{\circ},0^{\circ})$, and to $d_{x'^2-y'^2}$ at
$(75^{\circ},0^{\circ},0^{\circ})$.
The bottom left part of Figure \ref{Fig4} shows relative current contribution changes with respect to tip rotations by
$\psi_0$ around the sample $z$ direction starting from the $(45^{\circ},0^{\circ},0^{\circ})$ orientation.
We find that this type of rotation does not affect the dominant current contributions with $d_{y'z'}$ tip orbital
character. The biggest changes in other transitions are found for the sample $d_{yz}$ orbital, i.e., the contributions
in combination with the tip $d_{x'y'}$, $d_{y'z'}$, and $d_{x'z'}$ orbitals slightly increase, while the $d_{yz}-d_{3z'^2-r'^2}$
and $d_{yz}-d_{x'^2-y'^2}$ transitions show decreasing importance upon this kind of tip rotation.
Finally, by rotating the local tip coordinate system around the $z'$ axis with $\varphi_0$ starting from the
$(45^{\circ},0^{\circ},0^{\circ})$ orientation results in decreased $d_{y'z'}$ and increased $d_{x'z'}$ contributions in
combination with the sample $s$, $p_z$, and $d_{3z^2-r^2}$ orbitals. This is shown in the bottom right part of Figure \ref{Fig4}.
It is interesting to find that the $d_{3z^2-r^2}-s$ relative contribution increases by rotating $\varphi_0$. This, however,
does not mean an absolute increment of this current contribution since the tip $s$ state is insensitive to the rotation.

\subsection{Atomic contrast changes}
\label{sec_corrug}

On a constant-current ($I$=const) STM image, the sign change of the apparent height difference between the surface top
position ($z_T$) and hollow position ($z_H$),
\begin{equation}
\Delta z(I)=z_T(I)-z_H(I)
\end{equation}
is indicative for the corrugation inversion. [For the surface top (T) and hollow (H) positions, see Figure \ref{Fig2}.]
In the conventional understanding, $\Delta z(I)>0$ corresponds to a normal STM image, where the W atoms appear as protrusions,
and $\Delta z(I)<0$ to an anticorrugated image, where the W atoms show up as depressions \cite{palotas12orb,heinze98}.
We will demonstrate that this simple picture for the corrugation inversion does not hold considering the tip rotation effects on
the STM images. Instead, $\Delta z(I)$ gives information on the relative heights of the T and H positions only.
The tip rotations have more complex effects resulting in apparent atom positions that can be translated or
rotated with respect to the real atomic positions on the STM image.
Due to the monotonically decreasing character of the tunneling current with respect to the increasing tip-sample distance,
the current difference between tip positions above the T and H surface sites provides the same information on the
relative heights as $\Delta z(I)$ \cite{palotas12orb}.
The current difference at a tip-sample distance of $z$ and at bias voltage $V$ is defined as
\begin{equation}
\label{Eq_deltaI}
\Delta I(z,V)=I_T(z,V)-I_H(z,V).
\end{equation}
The $\Delta I(z,V)=0$ contour gives the $(z,V)$ combinations where the apparent heights of the surface T and H positions are
equal. The sign of $\Delta I(z,V)$ corresponds to the sign of $\Delta z(I(V))$ \cite{palotas12orb}.

\begin{figure*}
\includegraphics[width=1.00\textwidth,angle=0]{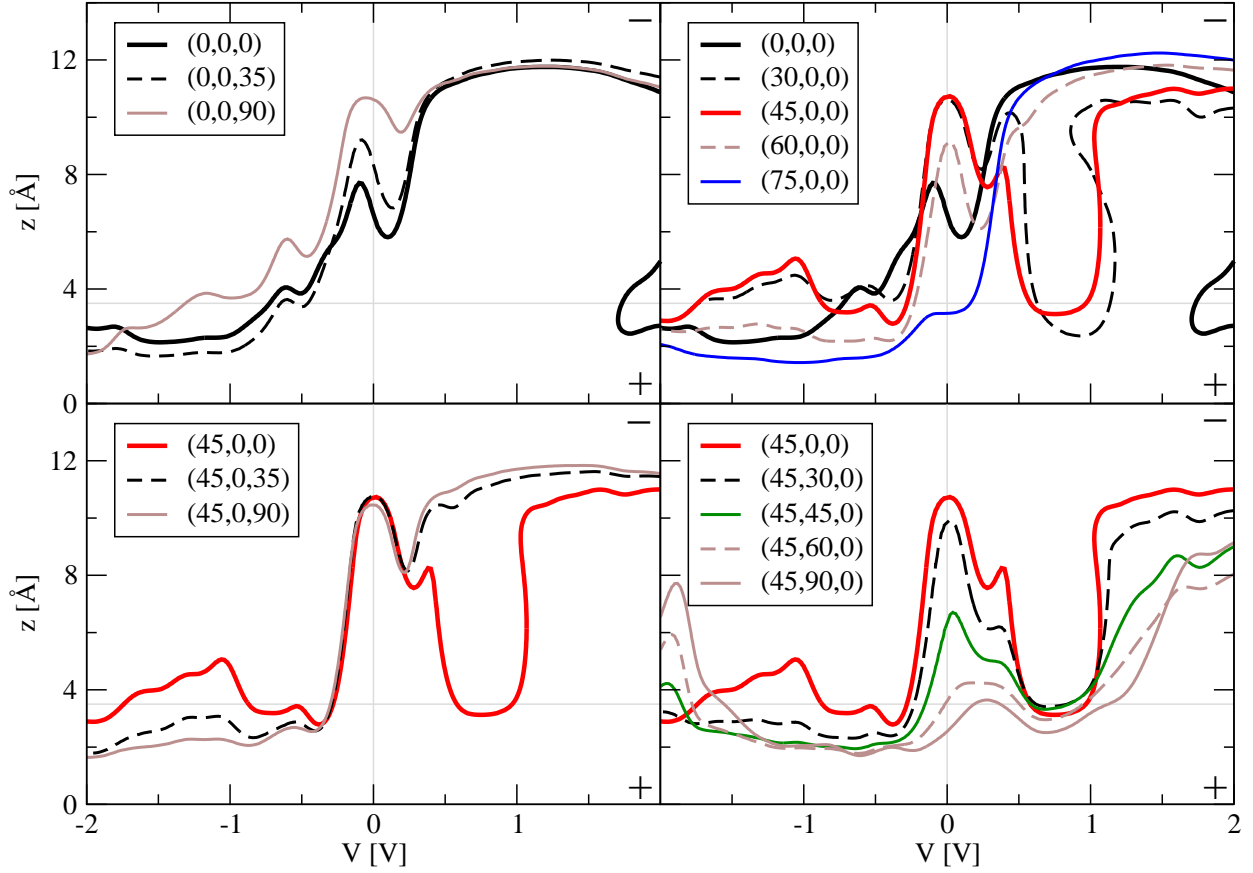}
\caption{\label{Fig5} (Color online) The $\Delta I(z,V)=I_T(z,V)-I_H(z,V)=0$ contours indicative for the relative apparent heights
of the surface top (T) and hollow (H) positions
[see Eq.(\ref{Eq_deltaI}), and its meaning in the text] calculated using the tungsten tip with different tip orientations
described by the Euler angles $(\vartheta_0,\varphi_0,\psi_0)$ given in degrees, see also Figure \ref{Fig1}.
The sign of $\Delta I(z,V)$ ($+$ or $-$) is explicitly shown at the corners on the right hand side of each part of the figure:
It is positive below the curves, and negative above them.
}
\end{figure*}

Figure \ref{Fig5} shows tip rotation effects on the $\Delta I(z,V)=0$ contours in the $[0$ \AA$,14$ \AA$]$ tip-sample distance
and [-2 V,+2 V] bias voltage range. Dotted vertical and horizontal lines denote the zero bias voltage and the limit of the
validity of any tunneling model, respectively. A pure tunneling model, e.g., the 3D WKB approach, is valid in the $z>3.5$ \AA\;
tip-sample distance range only.
We find that by rotating the tip with $\psi_0$ around the $z'=z$ axis (top left part of Figure \ref{Fig5}), the contours shift to
larger tip-sample distances close to zero bias, and their shapes remain qualitatively unchanged. It is interesting to see that the
$\Delta I(z,V)<0$ region found for the $(0^{\circ},0^{\circ},0^{\circ})$ tip orientation at around $z=3.5$ \AA\; close to +2 V
disappears by this type of tip rotation. The same finding is obtained in the top right part of Figure \ref{Fig5}, that
corresponds to rotations around the $x'=x$ axis with $\vartheta_0$. Here, the quality of the contours change considerably.
The $(30^{\circ},0^{\circ},0^{\circ})$ and $(45^{\circ},0^{\circ},0^{\circ})$ tip orientations result in $\Delta I(z,V)=0$
contours at enlarged tip-sample distances close to zero bias, and a $\Delta I(z,V)<0$ region opens at small tip-sample distances
between +0.5 V and +1 V bias voltages. By further rotation this region disappears, and concomitantly the contours shift to lower
tip-sample distances close to $V=0$ V. For the $(75^{\circ},0^{\circ},0^{\circ})$ tip orientation, we obtain $\Delta I(z,V)<0$ at
$z>3.5$ \AA\; around zero bias.
The bottom left part of Figure \ref{Fig5} shows the evolution of the $\Delta I(z,V)=0$ contours with respect to tip rotations by
$\psi_0$ around the sample $z$ direction starting from the $(45^{\circ},0^{\circ},0^{\circ})$ orientation.
The contours do not change considerably close to $V=0$ V, but the $\Delta I(z,V)<0$ region at small tip-sample distances between
+0.5 V and +1 V disappears.
Finally, the effect of the rotation of the local tip coordinate system around the $z'$ axis with $\varphi_0$ starting from the
$(45^{\circ},0^{\circ},0^{\circ})$ orientation is shown in the bottom right part of Figure \ref{Fig5}.
The contours are shifted to lower tip-sample distances close to zero bias and at high positive bias voltages, whereas the shift
is toward larger tip-sample distances at high negative bias. Moreover, this type of rotation does not affect the
presence of the $\Delta I(z,V)<0$ region at small tip-sample distances between +0.5 V and +1 V.

As it was suggested in Ref.\ \cite{palotas12orb}, particular tip nodal planes restrict the collection of surface atom
contributions to specific regions on the sample surface. By changing the tip-sample distance, the orientational overlaps between
the tip and sample orbitals change, and according to our model some localized orbitals gain more importance in the tunneling
contribution, see also Figure \ref{Fig4}. The complex tip-sample distance, bias-voltage, and tip-orientation dependent effect of
the real-space orbitals on the tunneling can be visualized as the zero contours of the current difference between tip positions
above the surface top and hollow sites, as shown in Figure \ref{Fig5}.

\begin{figure*}
\includegraphics[width=0.80\textwidth,angle=0]{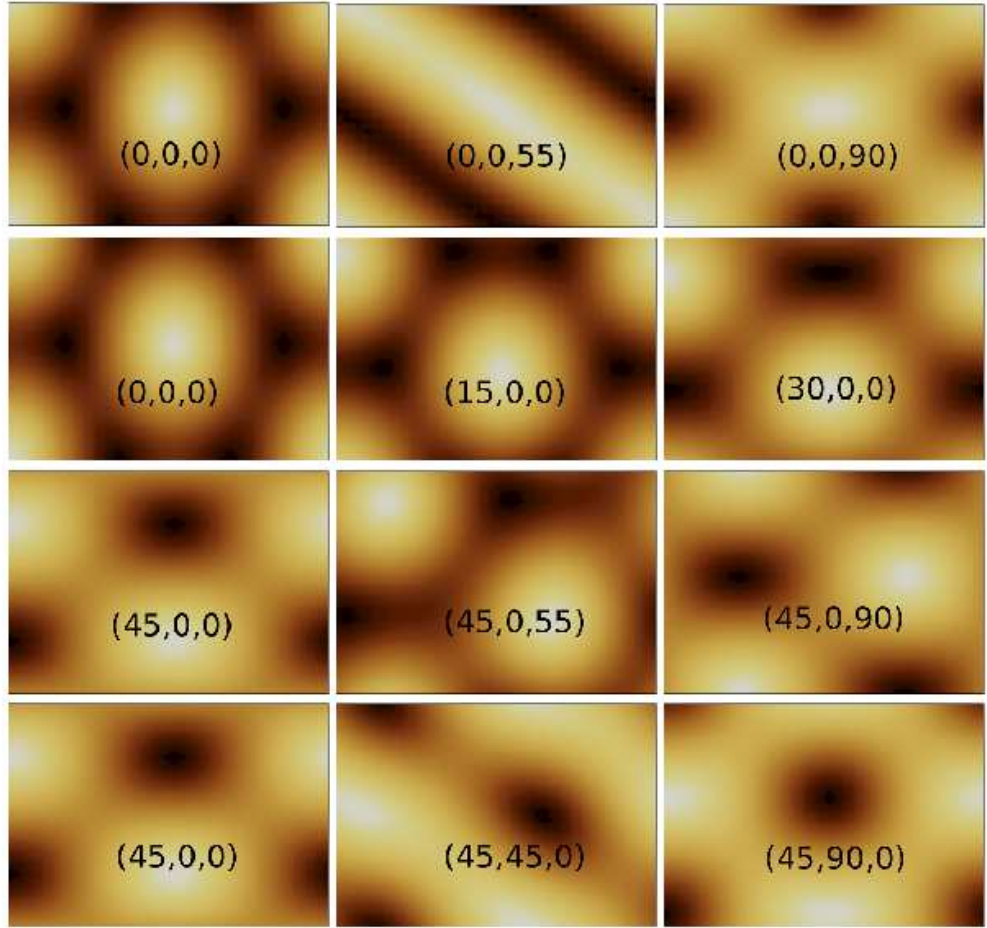}
\caption{\label{Fig6} (Color online) Tip orientation effect on the simulated STM images: Constant-current contours at $I$=6.3 nA
and $V$= -0.25 V bias voltage about $z$=4.5 \AA\; above the W(110) surface, using the tungsten tip with different orientations
described by the Euler angles $(\vartheta_0,\varphi_0,\psi_0)$ given in degrees, see also Figure \ref{Fig1}.
The scan area corresponds to the rectangle shown in Figure \ref{Fig2}. Light and dark areas denote larger and smaller
apparent heights, respectively.
}
\end{figure*}

To demonstrate the atomic contrast changes depending on the tip orientation $(\vartheta_0,\varphi_0,\psi_0)$ more apparently,
constant-current STM images are simulated. Selected results obtained at $I$=6.3 nA current and $V$= -0.25 V bias voltage are shown
in Figure \ref{Fig6}. The tunneling parameters correspond to tip-sample distances of about $z$=4.5 \AA, and the scan area
is the rectangular section shown in Figure \ref{Fig2}.
We find that by rotating the tip with $\psi_0$ around the $z'=z$ axis (top row of Figure \ref{Fig6}), the elongated feature
located on the W atoms initially in the $y$ direction is rotated. This results in a striped image for the
$(0^{\circ},0^{\circ},55^{\circ})$ tip orientation. The stripes with larger apparent height correspond to the atomic rows, and
they are oriented along the diagonal of the rectangle. Turning the tip to the $(0^{\circ},0^{\circ},90^{\circ})$ orientation, the
elongated feature located on the W atoms turns to the $x$ direction. The reason is the rearrangement of the importance of the
$d_{yz}-d_{y'z'}$ and $d_{xz}-d_{x'z'}$ transitions toward the $d_{yz}-d_{x'z'}$ and $d_{xz}-d_{y'z'}$ ones upon this type of
rotation, as shown in Figure \ref{Fig4}.
Tip rotation around the $x'=x$ axis with $\vartheta_0$ results in apparent atom positions shifted toward the bottom edge of the
image, i.e., toward the $-y$ direction. This effect is demonstrated for the set of images with $(0^{\circ},0^{\circ},0^{\circ})$
to $(45^{\circ},0^{\circ},0^{\circ})$ tip orientations (second row, and first image of the third row of Figure \ref{Fig6}).
During this rotation the dominant tip orbital character changes from $d_{3z'^2-r'^2}$ to $d_{y'z'}$, see Figure \ref{Fig4}.
The third row of Figure \ref{Fig6} shows the effect of tip rotations by $\psi_0$ around the sample $z$ direction starting from the
$(45^{\circ},0^{\circ},0^{\circ})$ orientation. We find that the apparent atom positions that were shifted away toward the $-y$
direction are now rotated on the images with respect to the $z$ axis centered on the real W atom positions. The STM image
corresponding to the $(45^{\circ},0^{\circ},55^{\circ})$ tip orientation shows apparent W atom positions shifted along the
diagonal of the rectangle with respect to the real atomic positions. Similarly, the $(45^{\circ},0^{\circ},90^{\circ})$ tip
orientation corresponds to apparent W atom positions shifted toward the $+x$ direction. As it was shown in Figure \ref{Fig4},
the tip $d_{y'z'}$ orbital is always dominant, and the biggest changes are found for the sample $d_{yz}$
orbital contributions upon this type of rotation.
The last row of Figure \ref{Fig6} considers the tip rotation around the $z'$ axis with $\varphi_0$ starting from the
$(45^{\circ},0^{\circ},0^{\circ})$ orientation. The obtained complex rearrangement of apparent atom positions on the STM images is
due to the changing effect of the $d_{y'z'}$ and $d_{x'z'}$ contributions of the tunneling tip, as demonstrated in Figure
\ref{Fig4}.

Thus, we highlighted the effect of a variety of tip orientations on the electron tunneling properties, particularly on the
occurrence of apparent atomic positions on the STM images of a W(110) surface.
Such tip orientation effects have to be considered at the evaluation of experimental STM images on other surfaces as well.
We suggest that the comparison of STM experiments to a large number of simulations with systematically varied tip orientations
could lead to a gain of more information on the local tip geometry and orientation. Combining tip rotations with different
crystallographic tip orientations and tip terminations
could enhance the agreement between experiment and theory considerably, as was demonstrated in Ref.\ \cite{hagelaar08}.
The 3D WKB atom-superposition theory \cite{palotas13fop} extended to include arbitrary tip orientations is a promising candidate
to be a powerful tool to perform the task of large scale simulations of the mentioned tip effects.

\section{Conclusions}
\label{sec_conc}

We extended the orbital-dependent electron tunneling model implemented within the 3D WKB atom-superposition approach for
simulating STM by including arbitrary tip orientations described by the Euler angles with respect to the sample coordinate system.
Applying our method, we highlighted the role of the real-space shape of the electron orbitals involved in the tunneling,
and analyzed the convergence and the orbital contributions of the tunneling current above the W(110) surface depending on the
orientation of a model tungsten tip. We found that tip rotations around the $z$ axis of the tip apex atom
do not change the dominating current contributions, while other rotations can change the tip character of the dominating
transitions. We also studied atomic contrast changes upon tip rotation. We found that the zero contours of the current difference
above the surface top and hollow positions have a complex tip-sample distance and bias-voltage dependence on the tip orientation.
The relative apparent heights of these two surface positions are directly related to the calculated current difference.
Simulating STM images at constant-current condition, we found that their quality depends very much on the tip orientation.
Some orientations result in protrusions on the images that do not occur above W atoms.
The presence of such apparent atom positions makes it difficult to identify the exact position of surface atoms. It is suggested
that this tip orientation effect should be considered at the evaluation of experimental STM images on other surfaces as well.
The presented computationally efficient tunneling model could prove to be useful for obtaining more information on the local tip
geometry and orientation by comparing STM experiments to a large number of simulations with systematically varied
tip orientations.
Extending this orbital-dependent tunneling model to magnetic junctions is expected to provide useful results about the interplay
of tip-orientation, real-space-orbital and spin-polarization effects in SP-STM and SP-STS experiments as well.

\section{Acknowledgments}

The authors thank W. A. Hofer and G. Teobaldi for useful discussions.
Financial support of the Magyary Foundation, EEA and Norway Grants, the Hungarian Scientific Research Fund (OTKA PD83353, K77771),
the Bolyai Research Grant of the Hungarian Academy of Sciences, and the New Sz\'echenyi Plan of Hungary
(Project ID: T\'AMOP-4.2.2.B-10/1--2010-0009) is gratefully acknowledged. Furthermore, partial usage of the
computing facilities of the Wigner Research Centre for Physics, and the BME HPC Cluster is kindly acknowledged.

\newpage

\begin{table}
\caption{Real-space orbitals, their definition from spherical harmonics
$Y_l^m(\vartheta^{(}$$'$$^{)},\varphi^{(}$$'$$^{)})$, and the angular dependence of their wave functions, i.e.,
real spherical harmonics $\chi_{\beta(\gamma)}(\vartheta^{(}$$'$$^{)},\varphi^{(}$$'$$^{)})$.
Note that $\vartheta^{(}$$'$$^{)}$ and $\varphi^{(}$$'$$^{)}$ are the usual polar and azimuthal angles, respectively,
in the spherical coordinate system centered on the corresponding sample (tip) atom.}
\label{Table1}
\begin{tabular}{|ccc|}
\hline
Orbital $\beta(\gamma)$ & Definition & $\chi_{\beta(\gamma)}(\vartheta^{(}$$'$$^{)},\varphi^{(}$$'$$^{)})$ \tabularnewline
\hline
\hline
$s$ & $Y_{0}^{0}$ & $1$ \tabularnewline
\hline
$p_{y^{(')}}$ & $Y_{1}^{1}-Y_{1}^{-1}$ & $\sin\vartheta^{(}$$'$$^{)}\sin\varphi^{(}$$'$$^{)}$\tabularnewline
$p_{z^{(')}}$ & $Y_{1}^{0}$ & $\cos\vartheta^{(}$$'$$^{)}$\tabularnewline
$p_{x^{(')}}$ & $Y_{1}^{1}+Y_{1}^{-1}$ & $\sin\vartheta^{(}$$'$$^{)}\cos\varphi^{(}$$'$$^{)}$\tabularnewline
\hline
$d_{x^{(')}y^{(')}}$ & $Y_{2}^{2}-Y_{2}^{-2}$ & $\sin^{2}\vartheta^{(}$$'$$^{)}\sin(2\varphi^{(}$$'$$^{)})$\tabularnewline
$d_{y^{(')}z^{(')}}$ & $Y_{2}^{1}-Y_{2}^{-1}$ & $\sin(2\vartheta^{(}$$'$$^{)})\sin\varphi^{(}$$'$$^{)}$\tabularnewline
$d_{3z^{(')2}-r^{(')2}}$ & $Y_{2}^{0}$ & $\frac{1}{2}(3\cos^{2}\vartheta^{(}$$'$$^{)}-1)$\tabularnewline
$d_{x^{(')}z^{(')}}$ & $Y_{2}^{1}+Y_{2}^{-1}$ & $\sin(2\vartheta^{(}$$'$$^{)})\cos\varphi^{(}$$'$$^{)}$\tabularnewline
$d_{x^{(')2}-y^{(')2}}$ & $Y_{2}^{2}+Y_{2}^{-2}$ & $\sin^{2}\vartheta^{(}$$'$$^{)}\cos(2\varphi^{(}$$'$$^{)})$\tabularnewline
\hline
\end{tabular}
\par
\end{table}

\end{document}